\documentclass[prl,twocolumn,preprintnumbers,amsmath,amssymb]{revtex4}
\usepackage{epsfig, graphicx}
\usepackage{hyperref}

\begin{document}

\title{Exact flow equation for composite operators}
\author{S. Floerchinger}
\author{C. Wetterich}
\affiliation{Institut f\"{u}r Theoretische Physik\\Universit\"at Heidelberg\\Philosophenweg 16, D-69120 Heidelberg}

\begin{abstract}
We propose an exact flow equation for composite operators and their correlation functions. This can be used for a scale-dependent partial bosonization or ``flowing bosonization'' of fermionic interactions, or for an effective change of degrees of freedom in dependence on the momentum scale. The flow keeps track of the scale dependent relation between effective composite fields and corresponding composite operators in terms of the fundamental fields.
\end{abstract}


\maketitle

The partial bosonization of fermionic interactions is a widely used technique for a theoretical description of complex many body systems. For example, a pointlike four-fermion interaction in the classical or microscopic action can be bosonized by using a Hubbard-Stratonovich transformation \cite{HS} in the functional integral. The resulting functional integral involves the original fermionic field as well as composite bosonic fields. After the transformation the fermionic part of the functional integral is Gaussian and can be performed explicitly. If the bosonic fluctuations are neglected, this yields mean field theory. 

A basic problem of mean field theory is the ``Fierz ambiguity''. Indeed, the choice of the bosonic fields is not unique, and different versions of the Hubbard-Stratonovich transformation often yield substantially different mean field results. Examples are the strong dependence of the phase diagram on the precise choice of the ``mean field'' for color superconductivity in QCD \cite{CS} or for the Hubbard model \cite{BaierBickCW}. The origin of this flaw is easily located -- it is the neglection of the important role of the bosonic fluctuations. Once the bosonic fluctuations are properly included in a given approximation, most of the unphysical dependence of the results on the choice of the mean field dissapears \cite{JaeckelCW}. Indeed, since the Hubbard-Stratonovich transformation is exact, any residual dependence on the choice of the mean field can be used as a test for the validity of approximations.

Partial bosonization of a fermionic interaction involves a second ambiguity that we may call the ``scale ambiguity''. The microscopic action has to be specified at some characteristic length scale $\Lambda^{-1}$. Changing $\Lambda$ changes the appropriate values of the couplings according to their renormalization flow. For example, a four-fermion coupling $\lambda_\psi(\Lambda)$ will depend on $\Lambda$ due to the fermionic fluctuations with momenta larger than $\Lambda$, which are included in the formulation of the microscopic theory at the scale $\Lambda$. Performing now partial bosonization via Hubbard-Stratonovich transformation at the scale $\Lambda$, one will find that the results of mean field theory depend on $\Lambda$ even if the running of $\lambda_\psi(\Lambda)$ has been taken into account. The reason for this dependence on the ``bosonization scale'' resides again in the neglection of the bosonic fluctuations with momenta smaller than $\Lambda$, while they are effectively taken into account at least partially for momenta larger than $\Lambda$  due to the running of $\lambda_\psi(\Lambda)$.

Exact flow equations for the ``average action'' or ``flowing action'' \cite{CWFloweq} are a convenient tool for dealing with these problems, since ``composite'' bosonic and ``fundamental'' fermionic fluctuations can be treated on the same footing \cite{CWBEA, GiesWetterich}. The problem of the Fierz and scale ambiguities finds a simple solution. Indeed, loops involving the bosonic fluctuations generate an effective four-fermion vertex at any length scale $k^{-1}>\Lambda^{-1}$, even if the four-fermion coupling $\lambda_\psi(\Lambda)$ has been eliminated by partial bosonization. If this is taken into account properly the dependence on the choice of the Hubbard-Stratonovich transformation is eliminated in principle and greatly reduced in practice already for simple approximative solutions \cite{JaeckelCW}. 

\begin{figure}[ht]
\centering
\includegraphics[width=0.25\textwidth]{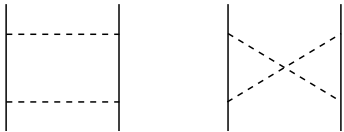}
\caption{Generation of four quark interactions by meson fluctuations. Fermion lines are solid, meson lines dashed.}
\label{fig:boxes}
\end{figure}

A good example is the Nambu-Jona-Lasino (NJL) model \cite{NJL} for pointlike strong interactions between the quarks. Partial bosonization eliminates the four-quark interaction $\lambda_\psi(\Lambda)$ in favor of a Yukawa interaction between the quarks and mesons. However, the meson fluctuations will induce again a four quark interaction, according to the diagram in Fig. \ref{fig:boxes}. It is possible to ``reabsorb'' this fluctuation-generated fermion interaction into a change of effective Yukawa couplings and meson masses, using a ``scale dependent partial bosonization'' or ``flowing bosonization'' \cite{GiesWetterich}. This solves the scale ambiguity since the scale $\Lambda$ chosen for the Hubbard-Stratonovich transformation no longer matters. (In praxis this holds only approximately due to truncations or other approximations.) The flowing bosonization constitutes in a sense a ``Hubbard-Stratonovich transformation at all scales''. Furthermore, the numerical contribution of the diagrams in Fig. \ref{fig:boxes} depends strongly on the choice of the mean field. This cancels the mean field ambiguity which results in different forms of the microscopic action, depending on the choice of the composite bosonic field.

There are various versions of flowing bosonization \cite{GiesWetterich, Pawlowski}. The versions used in practice so far catch the important qualitative ingredients and yield resonable results for practical calculations \cite{RebosonizationPractice}. The are, however, often related to exact equations that do not have the simple one-loop form of the original flow equation \cite{CWFloweq}, such that the quest for quantitative accuracy gets complicated. In this letter we present an exact flow equation for composite operators which yields a simple and exact flowing bosonization. Beyond the particular  setting with fermions and bosons the formalism can describe, in principle, arbitrary situations where composite degrees of freedom play an important role. The exactness and simplicity of the present version of flowing bosonization seems to be an ideal starting point for devising systematic approximations and expansions for partially bosonized theories.

\subsection{Scale dependent fields}
It may be instructive to start with the version of flowing bosonization which has been mainly used so far. It is based on a scale-dependent variable change for the flowing action \cite{GiesWetterich}. Consider the flowing action $\Gamma_k[\bar\psi,\bar \varphi]$ which depends on fundamental fields $\bar \psi$ (fermions or quarks) and composite fields $\bar\varphi$ (bosons or mesons). The dependence on the renormalization scale $k$ is given by the exact flow equation \cite{CWFloweq},
\begin{equation}
\partial_k \Gamma_k[\bar \psi,\bar \varphi] = \frac{1}{2} \text{STr} \left\{ (\Gamma_k^{(2)}+R_k)^{-1} \partial_k R_k
\label{eq:FloweqCW}
\right\}.
\end{equation}
On the right hand side of Eq.\ \eqref{eq:FloweqCW} appears $\Gamma_k^{(2)}$,  the second functional derivative of the flowing action with respect to the fields $\bar \psi$ and $\bar \varphi$. The equation has a one loop structure but is nevertheless exact. All orders in perturbation theory as well as non-perturbative effects are included. 

One possibility for a flowing bosonization scheme is to perform a $k$-dependent variable transformation on the field $\bar \varphi$. More specific, we write
\begin{equation}
\bar \varphi = \bar \varphi[\varphi]
\end{equation}
where $\bar\varphi[\varphi]$ is a $k$-dependent map that expresses the ``old fields'' $\bar \varphi$ in terms of the ``new fields'' $\varphi$. Now we consider the flow equation for $\Gamma_k[\varphi]=\Gamma_k[\bar \varphi[\varphi]]$ at fixed $\varphi$ (we suppress the argument $\bar \psi$)
\begin{equation}
\partial_k \Gamma_k{\big |}_{\varphi} = \partial_k \Gamma_k{\big |}_{\bar \varphi} - \int_q \frac{\delta \Gamma_k[\varphi]}{\delta \varphi} \partial_k \varphi {\big |}_{\bar \varphi}.
\label{eq:dernewfixed}
\end{equation}
The second term in Eq.\ \eqref{eq:dernewfixed} involves also a sum over possible internal degrees of freedom. It can be used to cancel certain terms generalized by the flow of the first term, as for example a pointlike four-fermion interaction. For this purpose the scale dependence $\partial_k \varphi|_{\bar \varphi}$ can be choosen arbitrarily. For example, a choice $\partial_k \varphi \sim \bar\psi \psi$, combined with $\partial \Gamma/\partial\varphi \sim h \bar \psi \psi$ as arising from a Yukawa coupling $\sim h \varphi \bar \psi\psi$ in the flowing action, generates a term in the flow $\sim (\bar \psi \psi)^2$. This can cancel a similar term generated by the first term in Eq.\ \eqref{eq:dernewfixed}, such that for $\partial_k \Gamma_k{|}_\varphi$ the sum vanishes. 

For the first term on the right hand side of Eq.\ \eqref{eq:dernewfixed} one can use the flow equation \eqref{eq:FloweqCW}. However, one should keep in mind that $\Gamma_k^{(2)}$ and the cutoff term $R_k$ are defined as derivatives of $\Gamma_k$ and $\Delta S_k$ with respect to the ``old fields'' $\bar \varphi$. A nonlinear coordinate change will lead to additional connection terms in the space of fields, since $\Gamma_k^{(2)}$ and $R_k$ are second derivatives and transform therefore as tensors of rank two \cite{CWFieldtransf}. These connection terms destroy the simple one-loop structure of the flow equation. They vanish for certain truncations and have been omitted for practical computations so far. We also note that the flowing action as a functional of the new fields $\Gamma_k[\varphi]$ differs in some properties from $\Gamma_k[\bar \varphi]$. For example, $\Gamma_k[\bar \varphi]$ always approaches a convex form for $k\to 0$ since it is then a Legendre transform. For $\Gamma_k[\varphi]$ this is not necessarily the case. 

\subsection{Scale-dependent bosonization}
In this note we aim for an exact flowing bosonization which keeps the simple one loop form of the original flow equation \eqref{eq:FloweqCW}. This will again modify the flow by additional ``tree contributions'' involving the first functional derivative of the flowing action. The structure differs, however, from the second term on the right hand side of Eq.\ \eqref{eq:dernewfixed}. The central idea is a scale-dependent Hubbard-Stratonovich transformation. 

Let us consider a scale-dependent Schwinger functional for a theory formulated in terms of the field $\tilde \psi$
\begin{equation}
e^{W_k[\eta]}=\int D\tilde\psi\, e^{-S_\psi[\tilde\psi]-\frac{1}{2}\tilde\psi_\alpha(R_k^\psi)_{\alpha\beta}\tilde\psi_\beta+\eta_\alpha\tilde\psi_\alpha}.
\label{eq:scaledepSF}
\end{equation}
We use here an abstract index notation where e.g. $\alpha$ stands for both continuous variables such as position or momentum and internal degrees of freedom. We now multiply the right hand side of Eq.\ \eqref{eq:scaledepSF} by a term that becomes for $R_k^\varphi=0$ and $j=0$ only a field independent constant. It has the form of the functional integral over the field $\tilde\varphi$ with a Gaussian weighting factor
\begin{equation}
\int D \tilde\varphi \, e^{-S_\text{pb}-\frac{1}{2}\tilde \varphi_\epsilon(R_k^\varphi)_{\epsilon\sigma}\tilde\varphi_\sigma+j_\epsilon\tilde\varphi_\epsilon},
\label{eq:Gaussianaddition}
\end{equation}
where
\begin{eqnarray}
\nonumber
S_\text{pb} &=& \frac{1}{2}\left(\tilde\varphi_\epsilon-\chi_\tau Q^{-1}_{\tau\epsilon}\right)Q_{\epsilon\sigma}(\tilde\varphi_\sigma-Q^{-1}_{\sigma\rho}\chi_\rho).\\
&=& \frac{1}{2}\left(\tilde\varphi-\chi Q^{-1}\right)Q(\tilde\varphi-Q^{-1}\chi),
\label{eq:Spb}
\end{eqnarray}
and $\chi$ depends on the ``fundamental field'' $\tilde \psi$. We will often supress the abstract index as in the last line of Eq.\ \eqref{eq:Spb}. We assume that the field $\tilde\varphi$ and the operator $\chi$ are bosonic. Without further loss of generality we can then also assume that $Q$ and $R_k^\varphi$ are $k$-dependent symmetric matrices.
As an example, we consider an operator $\chi$ which is quadratic in the original field $\tilde\psi$,
\begin{equation}
\chi_\epsilon = H_{\epsilon\alpha\beta}\tilde\psi_\alpha\tilde\psi_\beta.
\label{eq:chiYukawa}
\end{equation}

For $R_k^\varphi\neq0$, $j\neq0$ the multiplication of the integrand of Eq.\ \eqref{eq:scaledepSF} by the factor \eqref{eq:Gaussianaddition} defines a modified functional integral, for which the Schwinger functional reads
\begin{equation}
e^{W_k[\eta,j]} = \int D\tilde \psi\,D\tilde\varphi \, e^{-S_k[\tilde \psi, \tilde\varphi]+\eta\tilde\psi+j\tilde\varphi}
\label{eq:SFwithbosonfi}
\end{equation}
with
\begin{eqnarray}
\nonumber
S_k[\tilde\psi,\tilde\varphi] &=& S_\psi[\tilde\psi]+\frac{1}{2}\tilde\psi R_k^\psi\tilde \psi + \frac{1}{2}\tilde\varphi (Q+R_k^\varphi)\tilde\varphi\\
&&+ \frac{1}{2}\chi Q^{-1}\chi -\tilde\varphi \chi. 
\label{eq:actionferbos}
\end{eqnarray}
In the integration over $\tilde\varphi$, we can easily shift the variables to obtain
\begin{eqnarray}
\nonumber
e^{W_k[\eta,j]} &=& \int D \tilde\psi \, e^{-S_\psi[\tilde \psi]-\frac{1}{2}\tilde\psi R_k^\psi\tilde\psi+\eta \tilde\psi}\\
\nonumber
&& \times e^{\frac{1}{2}(j+\chi)(Q+R_k^\varphi)^{-1}(j+\chi)-\frac{1}{2}\chi Q^{-1}\chi}\\
&&\times \int D\tilde\varphi\, e^{-\frac{1}{2}\tilde\varphi(Q+R_k^\varphi)\tilde\varphi}.
\label{eq:Schingerfas}
\end{eqnarray}
The remaining integral over $\tilde\varphi$ gives only a ($k$-dependent) constant. For $R_k^\varphi=0$ and $j=0$ we note that $W_k[\eta,j]$ coincides with $W_k[\eta]$ in Eq.\ \eqref{eq:scaledepSF}.

We next derive identities for correlation functions of composite operators which follow from the equivalence of the equations\eqref{eq:SFwithbosonfi} and \eqref{eq:Schingerfas}. Taking the derivative with respect to $j$ we can calculate the expectation value for $\tilde \varphi$
\begin{eqnarray}
\nonumber
\varphi_\epsilon &=& \langle\tilde\varphi_\epsilon\rangle = \frac{\delta}{\delta j_\epsilon}  W_k[\eta,j]\\
&=& (Q+R_k^\varphi)^{-1}_{\epsilon\sigma}\,\left(j_\sigma+H_{\sigma\alpha\beta}\langle\tilde\psi_\alpha\tilde\psi_\beta\rangle\right).
\label{eq:varphiintermsofpsi}
\end{eqnarray}
This can also be written as
\begin{equation}
\langle\chi\rangle = Q\varphi - l
\label{eq:expvchi}
\end{equation}
with the modified source $l$
\begin{equation}
l_\epsilon = j_\epsilon-(R_k^\varphi)_{\epsilon\sigma}\varphi_\sigma.
\end{equation}
For the connected two-point function
\begin{equation}
(\delta_j\delta_j W_k)_{\epsilon\sigma}= \frac{\delta^2}{\delta j_\epsilon \delta j_\sigma} W_k =\langle\tilde\varphi_\epsilon\tilde\varphi_\sigma\rangle_c
\end{equation}
we obtain from Eq.\ \eqref{eq:Schingerfas} 
\begin{eqnarray}
\nonumber
&&(Q+R_k)(\delta_j\delta_j W_k)(Q+R_k) \\
\nonumber
&&= \langle(j+\chi)(j+\chi)\rangle -\langle(j+\chi)\rangle\langle(j+\chi)\rangle+(Q+R_k^\varphi)\\
&&= \langle\chi\chi\rangle-\langle\chi\rangle\langle\chi\rangle+(Q+R_k^\varphi)
\end{eqnarray}
or
\begin{eqnarray}
\nonumber
\langle\chi_\epsilon\chi_\sigma\rangle &=& \left[(Q+R_k^\varphi) (\delta_j\delta_j W_k)(Q+R_k^\varphi)\right]_{\epsilon\sigma} \\
&&+ (Q\varphi-l)_\epsilon(Q\varphi-l)_\sigma -(Q+R_k^\varphi)_{\epsilon\sigma}.
\label{eq:idk12}
\end{eqnarray}
Similarly, the derivative of Eq.\ \eqref{eq:expvchi} with respect to $j$ yields
\begin{eqnarray}
\nonumber
\langle\tilde\varphi_\epsilon\chi_\sigma\rangle = \langle\tilde\varphi_\epsilon\tilde\varphi_\tau\rangle (Q+R_k^\varphi)_{\tau\sigma}-\varphi_\epsilon j_\sigma -\delta_{\epsilon\sigma} \hspace{1.2cm}\\
= \varphi_\epsilon (Q\varphi)_\sigma + \left[(\delta_j\delta_j W_k)(Q+R_k^\varphi)\right]_{\epsilon\sigma}-\varphi_\epsilon l_\sigma-\delta_{\epsilon\sigma}. 
\label{eq:idk13}
\end{eqnarray}

\subsection{Flow equation}

We now turn to the scale-dependence of $W_k[\eta,j]$. In addition to $R_k^\psi$ and $R_k^\varphi$ also $Q$ and $H$ are $k$-dependent. For $H$ we assume
\begin{equation}
\partial_k H_{\epsilon\alpha\beta} = (\partial_k F_{\epsilon\rho}) H_{\rho\alpha\beta}
\end{equation}
where we take the dimensionless matrix $F$ to be symmetric for simplicity. For the operator $\chi$ this gives
\begin{equation}
\partial_k \chi_\epsilon = \partial_k H_{\epsilon\alpha\beta}\tilde\psi_\alpha\tilde\psi_\beta = \partial_k F_{\epsilon\rho} \chi_\rho.
\end{equation}
From Eqs.\ \eqref{eq:SFwithbosonfi} and \eqref{eq:actionferbos} we can derive (for fixed $\eta$, $j$)
\begin{eqnarray}
\nonumber
\partial_k W_k &=& -\frac{1}{2}\langle\tilde\psi(\partial_k R_k^\psi)\tilde\psi\rangle - \frac{1}{2}\langle\tilde\varphi(\partial_k R_k^\varphi+\partial_k Q)\tilde\varphi\rangle\\
\nonumber
&&-\frac{1}{2}\langle\chi\left(\partial_k Q^{-1}+Q^{-1}(\partial_k F)+(\partial_k F)Q^{-1}\right)\chi\rangle\\
&&+\langle\tilde\varphi(\partial_k F)\chi\rangle.
\end{eqnarray}
Now we insert Eqs.\ \eqref{eq:idk12} and \eqref{eq:idk13}
\begin{eqnarray}
\nonumber
\partial_k W_k &=& -\frac{1}{2}\psi (\partial_k R_k^\psi)\psi - \frac{1}{2}\varphi (\partial_k R_k^\varphi) \varphi\\
\nonumber
&&-\frac{1}{2} \text{STr}\,\{  (\delta_\eta\delta_\eta W_k)(\partial_k R_k^\psi) \}\\
\nonumber
&&- \frac{1}{2}\text{Tr}{\big \{} (\delta_j\delta_j W_k)(\partial_k R_k^\varphi){\big \}}\\
\nonumber
&&-\frac{1}{2}\text{Tr} {\big \{}{\big [}Q(\partial_k Q^{-1})R_k^\varphi+R_k^\varphi(\partial_k Q^{-1})Q\\
\nonumber
&&\,\,\,+R_k^\varphi(\partial_k Q^{-1})R_k^\varphi+ R_k^\varphi Q^{-1}(\partial_k F)(Q+R_k)\\
\nonumber
&&\,\,\,+(Q+R_k^\varphi)(\partial_kF)Q^{-1}R_k^\varphi{\big ]}(\delta_j\delta_j W_k){\big \}}\\
\nonumber
&&+\frac{1}{2}l\left[(\partial_k Q^{-1})Q+Q^{-1}(\partial_k F)Q\right]\varphi\\
\nonumber
&&+\frac{1}{2}\varphi \left[Q(\partial_k Q^{-1})+Q(\partial_k F)Q^{-1}\right]l\\
\nonumber
&&-\frac{1}{2}l \left[\partial_k Q^{-1}+Q^{-1}(\partial_k F)+(\partial_kF)Q^{-1}\right]l\\
\nonumber
&&+\frac{1}{2} \text{Tr} \left\{\left[\partial_k Q^{-1}+Q^{-1}(\partial_k F)+(\partial_k F)Q^{-1}\right]R_k^\varphi\right\}\\
&&+\frac{1}{2}\text{Tr} \left\{Q\partial_k Q^{-1}\right\}.
\label{eq:longflowW}
\end{eqnarray}
The supertrace $\text{STr}$ contains the appropriate minus sign in the case that $\tilde\psi_\alpha$ are fermionic Grassmann variables.

Equation \eqref{eq:longflowW} can be simplified substantially when we restrict the $k$-dependence of $F$ and $Q$ such that
\begin{equation}
\partial_k F = -Q(\partial_k Q^{-1}) = -(\partial_k Q^{-1})Q.
\label{eq:restrSQ}
\end{equation}
In fact, one can show that the freedom to choose $F$ and $Q$ independent from each other that is lost by this restriction, is equivalent to the freedom to make a linear change in the source $j$, or at a later stage of the flow equation in the expectation value $\varphi$. With the choice in Eq.\ \eqref{eq:restrSQ} we obtain
\begin{eqnarray}
\nonumber
\partial_k W_k &=& -\frac{1}{2}\psi (\partial_k R_k^\psi)\psi -\frac{1}{2}\varphi(\partial_k R_k^\varphi)\varphi\\
\nonumber
&&-\frac{1}{2}\text{STr}{\big \{}(\partial_k R_k^\psi)(\delta_\eta\delta_\eta W_k){\big \}}\\
\nonumber
&&-\frac{1}{2}\text{Tr}{\big \{} \left[\partial_k R_k^\varphi-R_k^\varphi(\partial_k Q^{-1})R_k^\varphi\right](\delta_j\delta_j W_k){\big \}}\\
&&+\frac{1}{2}l(\partial_k Q^{-1})l+\frac{1}{2}\text{Tr}{\{}\partial_kQ^{-1}(Q-R_k^\varphi){\}}.
\label{eq:shortflowW}
\end{eqnarray}
The last term is independent of the sources $\eta$ and $j$ and is therefore irrelevant for many purposes.

\subsection{Flowing action}
The average action or flowing action is defined by subtracting from the Legendre transform
\begin{equation}
\tilde\Gamma_k[\psi,\varphi] = \eta \psi + j \varphi - W_k[\eta,j]
\label{eq:Legendretransf}
\end{equation}
the cutoff terms
\begin{equation}
\Gamma_k[\psi,\varphi]=\tilde\Gamma_k[\psi,\varphi]-\frac{1}{2}\psi R_k^\psi\psi -\frac{1}{2}\varphi R_k^\varphi\varphi.
\label{eq:defflowingaction}
\end{equation}
As usual, the arguments of the effective action are given by
\begin{equation}
\psi_\alpha=\frac{\delta}{\delta \eta_\alpha}W_k \quad \text{and} \quad \varphi_\epsilon=\frac{\delta}{\delta j_\epsilon} W_k.
\end{equation}
By taking the derivative of Eq.\ \eqref{eq:defflowingaction} it follows
\begin{equation}
\frac{\delta}{\delta \psi_\alpha}\Gamma_k = \pm \eta_\alpha - (R_k^\psi)_{\alpha\beta} \psi_\beta,
\end{equation}
where the upper (lower) sign is for a bosonic (fermionic) field $\psi$. Similarly,
\begin{equation}
\frac{\delta}{\delta \varphi_\epsilon}\Gamma_k = j_\epsilon - (R_k^\varphi)_{\epsilon\sigma} \varphi_\sigma=l_\epsilon.
\end{equation}
In the matrix notation
\begin{eqnarray}
\nonumber
W_k^{(2)} &=& \begin{pmatrix}\delta_\eta\delta_\eta W_k, && \delta_\eta\delta_j W_k \\ \delta_j\delta_\eta W_k, && \delta_j\delta_j W_k\end{pmatrix},\\
\nonumber
\Gamma_k^{(2)} &=& \begin{pmatrix}\delta_\psi\delta_\psi \Gamma_k, && \delta_\psi\delta_\varphi \Gamma_k \\\delta_\varphi\delta_\psi \Gamma_k, && \delta_\varphi\delta_\varphi \Gamma_k\end{pmatrix},\\
R_k&=&\begin{pmatrix} R_k^\psi, &&  0 \\ 0,&& R_k^\varphi \end{pmatrix},
\end{eqnarray}
it is straight forward to establish
\begin{equation}
W_k^{(2)} \,\tilde\Gamma_k^{(2)} =1,\quad\quad W_k^{(2)} = (\Gamma_k^{(2)}+R_k)^{-1}.
\end{equation}

In order to derive the exact flow equation for the average action we use the identity
\begin{equation}
\partial_k \tilde \Gamma_k{\big |}_{\psi,\varphi} = -\partial_k W_k{\big |}_{\eta,j}.
\end{equation}
This yields our central result
\begin{eqnarray}
\nonumber
\partial_k \Gamma_k &=& \frac{1}{2}\text{STr}\, \left\{(\Gamma_k^{(2)}+R_k)^{-1}\left(\partial_k R_k-R_k(\partial_k Q^{-1})R_k\right)\right\}\\
&&-\frac{1}{2}\Gamma_k^{(1)} \left(\partial_k Q^{-1}\right)\Gamma_k^{(1)}+\gamma_k
\label{eq:flowequationGamma}
\end{eqnarray}
with
\begin{equation}
\gamma_k=-\frac{1}{2} \text{Tr}\left\{ (\partial_k Q^{-1})(Q-R_k)\right\}.
\end{equation}
As it should be, it reduces to the standard flow equation for a framework with fixed partial bosonization in the limit $\partial_k Q^{-1}=0$. The additional term is quadratic in the first derivative of $\Gamma_k$ with respect to $\varphi$ -- we recall that $\partial_k Q^{-1}$ has non-zero entries only in the $\varphi$-$\varphi$ block. Furthermore there is a field independent term $\gamma_k$ that can be neglected for many purposes.  

Before discussing practical consequences of a $k$-dependent partial bosonization a few remarks are in order.

(i) For $k\to 0$ the cutoffs $R_k^\psi$, $R_k^\varphi$ should vanish. This ensures that the correlation functions of the partially bosonized theory are simply related to the original correlation functions generated by $W_0[\eta]$, Eq.\ \eqref{eq:scaledepSF}, namely
\begin{eqnarray}
\nonumber
&&W_0[\eta, j] = \ln \left(\int D\tilde \psi \, e^{-S_\psi[\tilde \psi]+\eta\tilde\psi+jQ^{-1}\chi}\right)\\
\nonumber
&&\quad\quad\quad\quad\quad+\frac{1}{2}j\, Q^{-1} \,j+\text{const.},\\
&&W_0[\eta,j=0] = W_0[\eta] +\text{const.}
\end{eqnarray}
Knowledge of the dependence on $j$ permits the straightforward computation of correlation functions for composite operators $\chi$.
\newline

(ii) For solutions of the flow equation one needs a well known ``initial value'' which describes the microscopic physics. This can be achieved by letting the cutoffs $R_k^\psi$, $R_k^\varphi$ diverge for $k\to\Lambda$ (or $k\to\infty$). In this limit the functional integral in Eqs.\ \eqref{eq:SFwithbosonfi}, \eqref{eq:actionferbos} can be solved exactly and one finds
\begin{equation}
\Gamma_\Lambda[\psi,\varphi] = S_\psi[\psi] +\frac{1}{2}\varphi Q_\Lambda \varphi +\frac{1}{2}\chi[\psi]Q_\Lambda^{-1} \chi[\psi]-\varphi \chi[\psi].
\label{eq:averageactionmicrosc}
\end{equation}
This equals the ``classical action'' obtained from a Hubbard-Stratonovich transformation, with $\chi$ expressed in terms of $\psi$. 

(iii) In our derivation we did not use that $\chi$ is quadratic in $\psi$. We may therefore take for $\chi$ an arbitrary bosonic functional of $\psi$. It is straightforward to adapt our formalism such that also fermionic composite operators can be considered.

The flow equation \eqref{eq:flowequationGamma} has a simple structure of a one loop expression with a cutoff insertion -- $\text{STr}$ contains the appropriate integration over the loop momentum -- supplemented by a ``tree-contribution'' $\sim (\Gamma_k^{(1)})^2$. Nevertheless, it is an exact equation, containing all orders of perturbation theory as well as non-perturbative effects. The simple form of the tree contributions allows for easy implementations of a scale dependent partial bosonization. Furthermore, the flow Eq.\ \eqref{eq:flowequationGamma} is exact for an arbitrary choice of $Q$ and $R_k$. For a given approximation scheme for its solution, the residual dependence of the results on the choice of $Q$ and $R_k$ can therefore be used to judge the quality of the approximation. A particular simple form of the flow equation is obtained with the choice $R_k^\varphi=0$. In that case the terms involving $\varphi$ in the flowing action provide only a convenient way to parameterize the interactions of the original field $\psi$.

\subsection{Flowing bosonization of local four-fermion interaction}
Consider the simple case where the interaction terms in $S_\psi$ are given by a pointlike interaction of the form
\begin{equation}
S_\psi^{(\text{int})} = -\frac{1}{2}\int_x \lambda_\psi^{abcd}\bar \psi_a(x)\psi_b(x) \bar \psi_c(x) \psi_d(x)
\label{eq:localfourfermioninteraction}
\end{equation}
with
\begin{equation}
\lambda_\psi^{abcd} = (\tilde Q^{-1})_{ef} \tilde H_{eab} \tilde H_{fcd}.
\label{eq:BC}
\end{equation}
We now use a notation where the indices $a,b,c,...$ label internal degrees of freedom such as spin or flavor. 
We may define composite ``meson-operators''
\begin{equation}
\chi_e(x) = \tilde H_{eab} \bar \psi_a(x) \psi_b(x)
\end{equation}
such that
\begin{equation}
S_\psi^{(\text{int})} = -\frac{1}{2}\int_x \chi_e(x) \tilde Q_{ef}^{-1} \chi_f(x).
\end{equation}
Indeed, in a relativistic framework we may identify $\psi$ with quarks and $\bar \psi$ with antiquarks, such that the microscopic action describes the Nambu-Jona-Lasino model \cite{NJL}. (In a non-relativistic setting we may replace $\bar \psi$ by the Grassmann variable for holes $\psi^*$.) Choosing $Q_\Lambda^{-1}=\tilde Q^{-1}\,\delta(x-x^\prime)$ the initial action in Eq.\ \eqref{eq:averageactionmicrosc} describes a Yukawa interaction between quarks and mesons
\begin{equation}
\Gamma_\Lambda = S_{\psi,2} + \int_x \left(\frac{1}{2}\varphi_e \tilde Q_{ef} \varphi_f-\tilde H_{eab}\varphi_e\bar \psi_a\psi_b\right).
\label{eq:localquarkmeson}
\end{equation}
(We define $\bar \psi$ such that the action is invariant under hermitean conjugation for real $\varphi_e$.) The piece $S_{\psi,2}$ is quadratic in $\bar \psi,\psi$ and the four-fermion interaction has been transmuted to the Yukawa interaction by partial bosonization.

Following the flow for $k<\Lambda$ we may consider a truncation of the flowing action where $\Gamma_k$ keeps the form \eqref{eq:localquarkmeson}. Now $\tilde Q_{ef}$ and $\tilde H_{eab}$ are replaced by $k$-dependent quantities $\tilde q_{ef}$ and $\tilde h_{eab}$. They have to be distinguished from the quantities $\tilde Q$ and $\tilde H$ which relate the composite field $\varphi$ to the fundamental fields $\psi$ according to Eq.\ \eqref{eq:varphiintermsofpsi}. Also $\tilde Q$ and $\tilde H$ become $k$-dependent ``running couplings'' but they do not equal the running couplings $\tilde q$ and $\tilde h$ which appear in the flowing action. 

Even if we have removed the four fermion interaction at the microscopic scale $\Lambda$, it will typically be generated by the flow for $k<\Lambda$. One may therefore include in the truncation also a term $\Gamma_{k,4}$ which has the form \eqref{eq:localfourfermioninteraction}. Flow equations for the average action with quartic four-fermion interactions have been investigated in several interesting cases, ranging from QCD \cite{Ellwanger, MeggiolaroCW} to the Hubbard Model \cite{Metzner, SalmhoferHonerkamp}. Unfortunately, the contribution of the quartic interaction to the flow enhances substantially the complexity of the problem.

Let us assume that the dominant form of the four-fermion interaction can be accounted for by an effective meson-exchange interaction \eqref{eq:BC}, and that remaining interactions with a different structure can be neglected. The formalism of this note permits in this case for all scales $k$ a complete elimination of the four fermion interaction in favor of a modified running of the meson propagator and Yukawa interaction. This yields the considerable simplification that only interactions of the type in Eq.\ \eqref{eq:localquarkmeson} need to be retained in the truncation. Furthermore, the relevant meson physics is easily visible at all scales $k$ (in particular for $k\to0$) since it is all contained in the $\varphi$-dependent part of $\Gamma_k$ and not mixed with higher order fermion interactions.

The flow of the coupling $\lambda_\psi$ has now two contributions. First there is a ``direct contribution'' from the loop term in the flow equation, i.~e. the $\text{STr}$-expression in Eq.\ \eqref{eq:flowequationGamma}
\begin{equation}
k \partial_k \lambda_\psi^{abcd} = \beta_{\lambda_\psi,\text{dir}}^{abcd}.
\end{equation}
Second, the tree contribution yields
\begin{eqnarray}
\nonumber
\partial_k \Gamma_\text{tree} &=& -\frac{1}{2} \Gamma_k^{(1)}(\partial_k Q^{-1}) \Gamma_k^{(1)}\\
\nonumber
&=&-\frac{1}{2}\int_x \left(\tilde q_{ef}\varphi_f-\tilde h_{eab}\bar\psi_a \psi_b\right)(\partial_k \tilde Q^{-1})_{eg}\\
&&\times\left(\tilde q_{gh}\varphi_h - \tilde h_{gcd}\bar \psi_c \psi_d\right),
\end{eqnarray}
which reads explicitly
\begin{eqnarray}
\nonumber
\partial_k \Gamma_\text{tree}&=& \int_x {\bigg \{} -\frac{1}{2}\varphi_f (\tilde q \partial_k \tilde Q^{-1}\tilde q)_{fg}\varphi_g\\
\nonumber
&& +\frac{1}{2}\varphi_f \left[(\partial_k \tilde Q^{-1})\tilde q+\tilde q(\partial_k \tilde Q^{-1})\right]_{fg} \tilde h_{gab}\varphi_f \bar \psi_a \psi_b\\
&&-\frac{1}{2}(\partial_k \tilde Q^{-1})_{eg} \tilde h_{eab}\tilde h_{gcd}\bar \psi_a\psi_b \bar\psi_c\psi_d {\bigg \}}.
\label{eq:CC}
\end{eqnarray}
We can now choose $\partial_k \tilde Q^{-1}$ such that the tree contribution to the four fermion interaction in Eq.\ \eqref{eq:CC} precisely cancels the direct contribution
\begin{equation}
\beta_{\lambda_\psi,\text{dir}}^{abcd} + (k \partial_k \tilde Q^{-1})_{eg} \tilde h_{eab} \tilde h_{gcd} = 0.
\label{eq:CC01}
\end{equation}
This fixes the scale dependence of $\tilde Q$. In turn, the first two terms in Eq.\ \eqref{eq:CC} give additional contributions to the flow of the meson propagator and the Yukawa coupling which involve $\beta_{\lambda_\psi,\text{dir}}$. 

In order to faciliate comparison with earlier work \cite{GiesWetterich} we concentrate on the one-component NJL-model, where the complex meson field $\varphi=(\varphi_1+i \varphi_2)/\sqrt{2}$ is expressed in terms of two real scalar fields $\varphi_1$, $\varphi_2$. There is a single Yukawa coupling $\bar h$ according to
\begin{eqnarray}
\nonumber
\tilde h_{1ab}\bar \psi_a \psi_b &=& \frac{\bar h}{\sqrt{2}}\left(\bar \psi_L\psi_R-\bar\psi_R\psi_L\right),\\
\tilde h_{2ab}\bar \psi_a \psi_b &=& -\frac{i \bar h}{\sqrt{2}}\left(\bar \psi_L\psi_R+\bar\psi_R\psi_L\right)
\end{eqnarray}
and we define a ``compositeness scale'' $m_c$ by
\begin{equation}
\tilde Q_{ef} = m_c^2 \delta_{ef}.
\end{equation}
For $k=\Lambda$ we identify $m_c^2=\bar m_\Lambda^2$ such that the microscopic action \eqref{eq:localquarkmeson} reads
\begin{equation}
\Gamma_\Lambda = S_{\psi,2} + \int_x\left(\bar m_\Lambda^2 \varphi^*\varphi+\bar h_\Lambda \varphi \bar \psi_R \psi_L-\bar h_\Lambda \varphi^* \bar\psi_L \psi_R\right).
\end{equation}
The corresponding four fermion interaction Eq.\ \eqref{eq:localfourfermioninteraction} becomes
\begin{equation}
S_\psi^{(\text{int})} = \frac{\bar h^2}{m_c^2}\int_x (\bar \psi_R \psi_L)(\bar \psi_L\psi_R).
\end{equation}
In the standard convention for the NJL-coupling $\lambda_\sigma$ this corresponds to
\begin{equation}
\lambda_\sigma = \frac{\bar h^2}{2 m_c^2}.
\end{equation}

Assume now that the direct or loop contribution to the flow of $\lambda_\sigma$ has been computed,
\begin{equation}
k \partial_k \lambda_\sigma = \beta_{\lambda_\sigma,\text{dir}}=\beta_\sigma.
\label{eq:G6}
\end{equation}
Eq.\ \eqref{eq:CC01} then translates to a flow equation for $m_c$
\begin{equation}
k\partial_k m_c^{-2} = -2\beta_\sigma/\bar h^2. 
\end{equation}
If we denote 
\begin{equation}
\tilde q_{ef} = \bar m_\varphi^2 \delta_{ef}
\label{eq:qglmphi}
\end{equation}
the flow of the scalar mass term $\bar m_\varphi^2$ receives a contribution $\sim \beta_\sigma$ according to Eq.\ \eqref{eq:CC}
\begin{equation}
k \partial_k \bar m_\varphi^2 = \beta_{\bar m_\varphi^2,\text{dir}} + 2 \beta_\sigma \frac{\bar m_\varphi^4}{\bar h^2}.
\label{eq:mphi2mod}
\end{equation}
Here $\beta_{\bar m_\varphi^2,\text{dir}}$ accounts for the running due to the loop contribution. If $\beta_\sigma$ is negative, corresponding to a growing $\lambda_\sigma$ for decreasing $k$ according to Eq.\ \eqref{eq:G6}, the ``running bosonization'' of the four-fermion vertex tends to increase $\bar m_\varphi^2$ for decreasing $k$. On the other hand, we also observe a contribution to the running Yukawa coupling
\begin{equation}
k \partial_k \bar h = \beta_{\bar h,\text{dir}}+ 2 \beta_\sigma \frac{\bar m_\varphi^2}{\bar h}
\label{eq:hmod}
\end{equation}
which enhances $\bar h$ for decreasing $k$. The additional contribution to the ratio $\bar h^2/(2 \bar m_\varphi^2)$
\begin{equation}
\Delta \left(k \frac{\partial}{\partial k}\frac{\bar h^2}{2 \bar m_\varphi^2}\right) = \beta_\sigma
\label{eq:Deltalambda}
\end{equation}
shows that the effective coupling is enhanced by the running bosonization as $k$ decreases.

The modified flow of $\bar m_\varphi^2$ and $\bar h$, as given by Eqs.\ \eqref{eq:mphi2mod}, \eqref{eq:hmod} specifies how a momentum independent part of the flowing four fermion vertex $\lambda_\psi$ can be absorbed into effective bosonic interactions at every scale $k$. One could further improve and also absorb a momentum dependent part of $\lambda_\psi$. Any contribution to $\lambda_\psi$ which only depends on the momentum exchange in the meson channel can be absorbed by generalizing Eq.\ \eqref{eq:CC01} to a momentum dependent $\tilde Q$. The additional piece in the flow of the meson propagator and Yukawa coupling become then momentum dependent. We also note that an extended truncation of $\Gamma_k$ with a momentum dependent meson propagator results in momentum dependent modifications even if only the momentum independent part of $\lambda_\psi$ is bosonized. In Eq.\ \eqref{eq:CC} we then have to replace $\tilde q$ by a momentum dependent function. 

\subsection{Comparison to scale dependent fields}

In ref.\ \cite{GiesWetterich} the flow equation \eqref{eq:dernewfixed} was employed for an investigation of the NJL model. Here the scale dependence of the field $\varphi$ is chosen to be of the form $\partial_k \varphi = -\bar \psi_L\psi_R \partial_k \alpha$ and $\partial_k \varphi^* = \bar \psi_R\psi_L \partial_k \alpha$ with some function $\alpha$ that is choosen for convenience. As discussed in \cite{GiesWetterich} this allows in the limit of pointlike interactions for a scale dependent bosonization with vanishing coupling $\lambda_\sigma$ on all scales. Let us discuss how the results of \cite{GiesWetterich} are modified if the flow equation \eqref{eq:flowequationGamma} is used instead of Eq.\ \eqref{eq:dernewfixed}. We neglect in this discussion the connection terms that appear in Eq.\ \eqref{eq:dernewfixed}. 

First, the term
\begin{equation}
-\frac{1}{2}\text{Tr} \left\{(\Gamma_k^{(2)}+R_k)\,R_k(\partial_k Q^{-1})R_k\right\}
\label{eq:nB}
\end{equation}
in Eq.\ \eqref{eq:flowequationGamma} has no correspondence in Eq.\ \eqref{eq:dernewfixed}. However, this term is only a correction to the loop term in the usual flow equation
\begin{equation}
\frac{1}{2}\text{STr} (\Gamma_k^{(2)}+R_k)^{-1} \partial_k R_k.
\end{equation}
The term \eqref{eq:nB} modifies the contribution of loops that involve the composite scalar field $\varphi$ to the flow of $\Gamma_k$. Loops involving only the original field $\psi$ or other fields as the gauge field $A_\mu$ in \cite{GiesWetterich} are not modified. Even for those loop expressions where \eqref{eq:nB} leads to modifications, it does not change the qualitative structure but can be seen as a quantitative correction to the cutoff derivative
\begin{equation}
\partial_k R_k \to \partial_k R_k - R_k (\partial_k Q^{-1}) R_k.
\end{equation}

More important, the two approaches differ in the additional ``tree-level term'', i.~e.\
\begin{equation}
-\frac{1}{2} \Gamma_k^{(1)} (\partial_k Q^{-1}) \Gamma_k^{(1)}
\end{equation}
in Eq.\ \eqref{eq:flowequationGamma}, as compared to the term $\sim \frac{\delta \Gamma_k}{\delta \varphi}$ in Eq.\ \eqref{eq:dernewfixed}. For a simple truncation as the one employed here and in \cite{GiesWetterich}, the flow equation \eqref{eq:dernewfixed} yields 
\begin{eqnarray}
\nonumber
k \partial_k \bar m_\varphi^2 &=& \beta_{\bar m_\varphi^2,\text{dir}},\\
k \partial_k \bar h &=& \beta_{\bar h,\text{dir}} +   \beta_\sigma \frac{\bar m_\varphi^2}{\bar h}.
\label{eq:flowmhGW}
\end{eqnarray}
Comparison with Eqs.\ \eqref{eq:mphi2mod}, \eqref{eq:hmod} shows that $\bar m_\varphi^2$ remains now unchanged, while the flow of $\bar h$ gets only half the contributions as compared to Eq.\ \eqref{eq:hmod}. The flow \eqref{eq:flowmhGW} of $\bar h^2/(2\bar m_\varphi^2)$ is the same however. As long as we remain within an approximation of pointlike effective four fermion interactions the two methods give the same result. The differences become manifest beyond the pointlike interaction. This is already apparent if a momentum dependence is included in the effective meson propagator.

In conclusion, the scale dependent bosonization offers the advantage of a simple exact flow equation which keeps the one loop structure. It permits to transfer part of the interactions of the fundamental fields (fermions) to a correction to the effective interaction of composite fields (bosons). Of course, one can also solve the flow equations directly for the fermionic vertex $\lambda_\psi$ without ever proceeding to bosonization. Interesting results have been obtained by taking the momentum depdendence of $\lambda_\psi$ into account \cite{Ellwanger, Metzner, SalmhoferHonerkamp}. Partial bosonization can be used in addition to the flow of $\lambda_\psi$ by treating the dominant part more accurately. In a first approximation it can even replace $\lambda_\psi$, by neglecting those parts that cannot be absorbed by a given choice of the flowing bosonization. The advantage of bosonization is the focus on the most relevant degrees of freedom. 
This is most important in case of spontaneous symmetry breaking by composite fields.

\end{document}